\documentclass[twocolumn,twoside,slac_two]{revtex4}
\usepackage{graphicx}
\usepackage{fancyhdr}
\begin{document}

\title{HESS J1507-622: an unique unidentified source off the Galactic Plane}

%

\author{O. Tibolla, W. Domainko, W. Hofmann}
\affiliation{Max-Planck-Institut f\"ur Kernphysik, Heidelberg, P.O. Box 103980, D69029 Heidelberg, Germany}
\author{O. de Jager}
\affiliation{Unit for Space Physics, North-West University, Potchefstroom 2520, South Africa}
\author{S. Kaufmann}
\affiliation{Landessternwarte, Universit\"at Heidelberg, K\"onigstuhl, D 69117 Heidelberg, Germany}
\author{N. Komin}
\affiliation{CEA, Irfu, SPP, Centre de Saclay, F-91191 Gif-sur-Yvette, France}
\author{K. Kosack}
\affiliation{CEA, Irfu, SAP, Centre de Saclay, F-91191 Gif-sur-Yvette, France}
\author{on behalf of the H.E.S.S. collaboration}
\affiliation{..\\}
\begin{abstract}
Galactic very high energy (VHE, $> 10^{11}$ eV) gamma ray sources in the inner Galaxy H.E.S.S. survey tend to cluster within 1 degree in latitude
around the Galactic plane. HESS J1507-622 instead is unique, since it is located at latitude of $\sim$3.5 degrees.
HESS J1507-622 is slightly extended over the PSF of the instrument and hence its Galactic origin is clear. The search for counterparts in other
wavelength regimes (radio, infrared and X-rays) failed to show any plausible counterparts; and given its position off the Galactic plane and hence the absorption almost one order of magnitude lower, it is very surprising to not see any counterparts especially at X-rays wavelengths (by ROSAT, XMM Newton and Chandra). Its latitude implies that it is either rather close, within about 1 kpc, or is located well off the Galactic plane. And also the models reflect the uniqueness of this object: a leptonic PWN scenario would place this source due to its quite small extension to multi-kpc distance whereas a hadronic scenario would preferentially locate this object at distances of $<$ 1 kpc where the density of target material is higher.
\end{abstract}
\maketitle

\thispagestyle{fancy}


\section{Introduction}
Very high energy (VHE, $> 10^{11}$ eV) particles can be traced within our Galaxy by a combination of non-thermal X-ray emission and VHE gamma-ray emission via leptonic (i.e. Inverse Compton scattering of electrons, Bremsstrahlung and synchrotron radiation) or hadronic (i.e. the decay of charged and neutral pions, due to interactions of energetic hadrons) processes.

The High Energy Stereoscopic System (H.E.S.S.) detects VHE $\gamma$-rays above an energy threshold of $\sim$100 GeV and up to $\sim$100 TeV with a typical energy resolution of 15\% per photon. The angular resolution is $\sim 0.1^{\circ}$ per event, allowing a positional error better than 40'' for a point source detected with a statistical significance of 6 $\sigma$. The H.E.S.S. field of view is almost $5^{\circ}$ in diameter with a point source sensitivity of $<2.0 \times 10^{-13}$ ergs cm$^{-2}$ s$^{-1}$ ($\sim$1\% of the Crab Nebula) for a  5 $\sigma$ detection in 25 hours of observations \cite{1}.

\section{H.E.S.S. observations and analysis}
During the extension of the H.E.S.S. survey of the Galactic plane, performed in 2006/2007 \cite{ryan}, there were indications of a source at the position of HESS J1507-622; follow-up dedicated observations were performed for 9.7 hours (21 runs of ~28 minutes each) in 2007 and in 2008.

\begin{figure}[t]
   \centering
\includegraphics[width=\columnwidth]{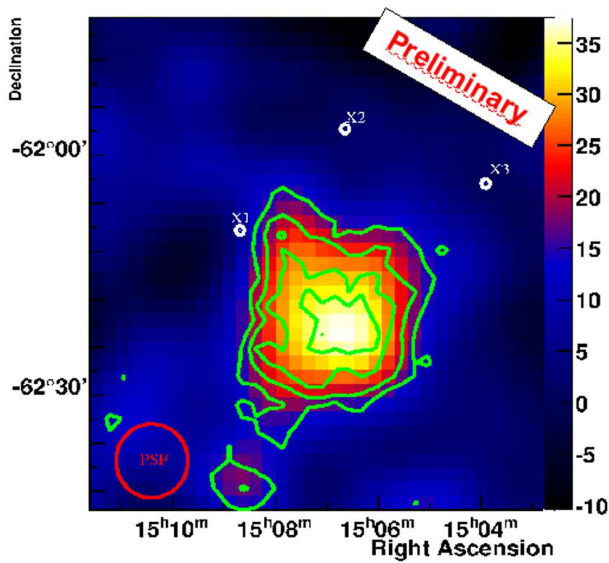}
  	\caption{Excess count map (smoothed with Gaussian of $0.07^{\circ}$ radius) of HESS J1507-622. The green contours show 3$\sigma$, 4$\sigma$, 5$\sigma$, 6$\sigma$ of significance, respectively, for an integration radius of $0.12^{\circ}$. The white circles represent the position of three faint RASS (\cite{faintRASS}) sources: X1 indicates 1RXS J150841.2-621006, X2 indicates 1RXS J150639.1-615704 and X3 indicates 1RXS J150354.7-620408. The 68\% containment radius for the Point Spread Function (PSF) of the H.E.S.S. instrument for these observations is superimposed in magenta.}
         \label{1507}
   \end{figure}

HESS J1507-622 is visible in Figure \ref{1507}.
The peak significance is 9.3 $\sigma$ for the 9.7 hours of dedicated observations (employing a $0.22^{\circ}$ oversampling radius, which is the standard radius used in source searches in the H.E.S.S. Galactic plane survey \cite{2}). The figure shows the uncorrelated excess count map (smoothed with Gaussian of $0.07^{\circ}$) for the dedicated observations, using hard cuts \cite{1} and Ring \cite{berge} background method. A two
dimensional Gaussian fit shows that the best fit position is at RA = $226.72^{\circ} \pm 0.05^{\circ}$, Dec = $-62.35^{\circ} \pm 0.03^{\circ}$, and the source is slightly extended with intrinsic size (not including the PSF) of $0.15^{\circ} \pm 0.02^{\circ}$ radius. The green contours show 3 $\sigma$, 4 $\sigma$, 5 $\sigma$ and 6 $\sigma$ of significance, respectively; the peak significance for this smaller ($0.12^{\circ}$) correlation radius is 7.0 $\sigma$.

\begin{figure}[t]
   \centering
   \includegraphics[width=\columnwidth]{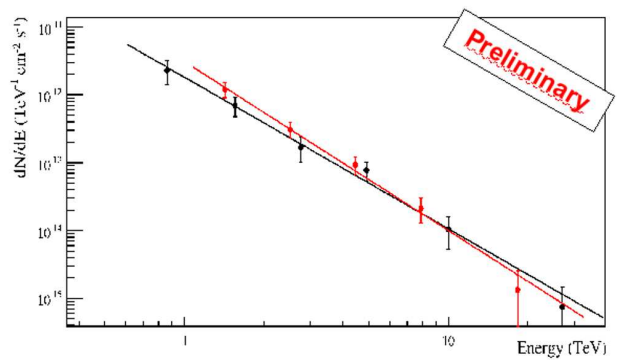}
      \caption{The VHE spectrum observed from HESS J1507-622. The black line represents the best $\chi^2$ fit of a power law to observed data using the \emph{standard cuts}, whereas the red line shows the spectral result using \emph{hard cuts} \cite{1}.
              }
         \label{spec}
   \end{figure}

Using standard cuts \cite{1} and hence a lower energy threshold ($\sim$ 500 GeV), the observed spectrum can be well fitted with a power-law dN/dE = k (E/1 TeV)$^{- \Gamma}$ with photon index $\Gamma = 2.24 \pm 0.16_{stat} \pm 0.20_{sys}$ and a flux normalization $k = (1.8 \pm 0.4) \times 10^{-12}$ TeV$^{-1}$ cm$^{-2}$ s$^{-1}$; hence the integral flux ($>$ 1 TeV) is $(1.5 \pm 0.4_{stat} \pm 0.3_{sys}) \times 10^{-12}$ cm$^{-2}$ s$^{-1}$. Using hard cuts \cite{1} (energy threshold $\sim$ 1 TeV) and hence a better gamma-hadron separation, the observed spectrum fits well with a power-law with photon index $ \Gamma = 2.49 \pm 0.18_{stat} \pm 0.20_{sys}$ and a flux normalization $k = (3.1 \pm 0.8) \times 10^{-12}$ TeV$^{-1}$ cm$^{-2}$ s$^{-1}$; hence the integral flux ($>$ 1 TeV) is $(2.1 \pm 0.6_{stat} \pm 0.4_{sys}) \times 10^{-12}$ cm$^{-2}$ s$^{-1}$. The data points are compatible; the difference in flux arises from slope difference and from extrapolation to 1 TeV. The spectra are visible in Figure \ref{spec}.

\section{Radio and infrared observations}

At radio and infrared wavelengths, HESS J1507-622 is too far offset from the Galactic Plane to be covered by the Southern Galactic Plane Survey \cite{sgps} and by Spitzer GLIMPSE \cite{spitzer}. This region of the sky had been covered by the Midcourse Space Experiment (MSX) in all its
wave bands \cite{msx} and by the MOLONGLO Galactic plane survey \cite{molonglo}, but without evidence for any plausible counterpart. 

\begin{figure}[t]
   \centering
   \includegraphics[width=\columnwidth]{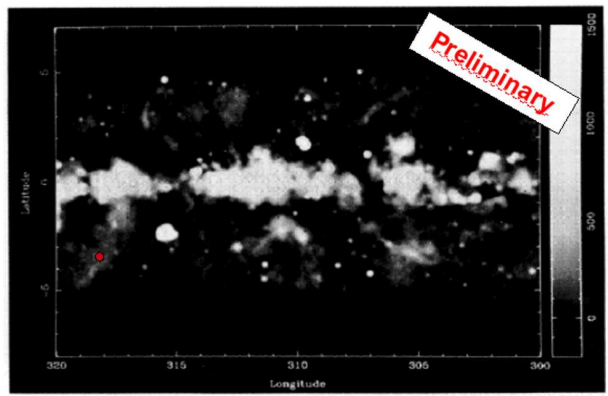}
      \caption{ The 2.4 GHz Radio map \cite{duncan}; HESS J1507-622 is overlapped in red.
              }
         \label{rad}
   \end{figure}

HESS J1507-622 is located on a radio emission filament, visible in Figure \ref{rad}, at 2.4 GHz, which was tentatively considered a part of a very large ($\sim 15^{\circ}$ in diameter) and nearby candidate SNR \cite{duncan}.
In the new complete CO survey \cite{dame}, the H.E.S.S. source lies near the edge of a large ($ \sim 5^{\circ} \times \sim 2^{\circ}$) nearby CO molecular cloud; the peak velocity of this cloud, around -5 km/s, would most likely place it at the near distance of $\sim$400 pc. The substantial difference in extension and, in the case of the CO molecular cloud, the offset of $\sim 1^{\circ}$ from the H.E.S.S. source centroid, suggest no obvious scenario for an association.

\section{X-ray observations}

\begin{figure}[t]
   \centering
   \includegraphics[width=\columnwidth]{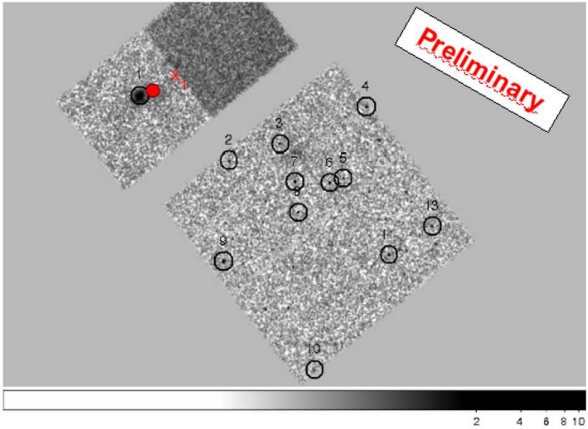}
      \caption{ Background subtracted and smoothed count map of the Chandra observations of HESS J1507-622 in the range 0.3-8 keV. In red, labeled X1, the faint ROSAT source 1RXS J150841.2-621006.
              }
         \label{ch}
   \end{figure}

At X-rays, the source was observed with XMM-Newton in the AO-7 (proposal number: 055631; PI: O. Tibolla); but unfortunately the presence of a huge soft proton flare affected dramatically the 25 ks observation, leading us to only 0.8 ks of GTI for the PN detector, 8.0 ks for MOS1 and 9.2 ks for MOS2. The GTI data have been analyzed with SAS 9.0, discovering one point-like source (source 7 in Figure \ref{ch}), confirmed by Chandra. 

The source was observed with Chandra in the AO-10 (proposal number: 10400599; PI: O. Tibolla): Figure \ref{ch} shows the count map between 0.3 and 8 keV as seen by Chandra.
Chandra data have been analyzed using CIAO 4.1: 12 point-like sources (mainly identified as stars) and one extended source have been discovered.
Hints for diffuse emission are visible between source 3 and source 7 in Figure \ref{ch}, and are still under evaluation.

\section{Discussion and Conclusion}

Given past experience, deeper multi-wavelength data has in most cases revealed a possible counterpart, usually PWN or SNR, however HESS J1507-622 is so far unidentified. The offset can be either interpreted that HESS J1507-622 is a nearby object and the large angular offset does not correspond to a large physical distance from the disk or that it is truly distant.

Both leptonic and hadronic scenarios are currently under investigations. At a location away from the Galactic disk, the density of target material is quite low, which would make a hadronic scenario for HESS J1507-622 less favorable. On the other hand, the Cosmic Microwave Background (CMB), as target photon field for gamma-ray production, is available everywhere and would point towards a leptonic gamma-ray emission mechanism; in particular the case of an ancient PWN \cite{ancientPWN} is under evaluation. 
However, if confirmed by deeper observations, the lack of emission in other wavebands could imply a hadronic origin of HESS J1507-622. The difference in model leads also to a difference in the distance: a leptonic PWN scenario would place this source due to its quite small extension to multi-kpc distance whereas a hadronic scenario would preferentially locate this object at distances of $<$ 1 kpc where the density of target material is higher.

\section{Acknowledgements}
The support of the Namibian authorities and of the University of Namibia in facilitating the construction and operation of HESS is gratefully acknowledged, as is the support by the German Ministry  for Education and Research (BMBF), the Max Planck Society, the French Ministry for Research, the CNRS-IN2P3 and the Astroparticle Interdisciplinary Programme of the CNRS, the U.K. Science and Technology Facilities Council (STFC), the IPNP of the Charles University, the Polish Ministry of Science and Higher Education, the South African Department of Science and Technology and National Research Foundation, and by the University of Namibia. We appreciate the excellent work of the technical support staff in Berlin, Durham, Hamburg, Heidelberg, Palaiseau, Paris, Saclay, and in Namibia in the construction and operation of the equipment.











\end{document}